\documentclass[12pt,preprint]{aastex}
\tighten

\usepackage{natbib} 

\newcommand\ha{{H$\alpha$}}

\newcommand{\msun}{M_\odot}

\newcommand\masy{\:\rm{\,mas\:yr^{-1}}}

\newcommand\perpix{\:{\rm pixel}^{-1}}

\newcommand\spitzer{{\em Spitzer}}

\slugcomment{Accepted for publication in  {\it The Astrophysical Journal}}

\shorttitle{IR Emission from SN\,1006}
\shortauthors{Winkler et al.}

\begin{document}
\title{The First Reported Infrared Emission from the SN\,1006 Remnant}
\author{P. Frank Winkler\altaffilmark{1}, Brian J. Williams\altaffilmark{2}, William P. Blair\altaffilmark{3}
Kazimierz J. Borkowski\altaffilmark{4}, Parviz Ghavamian\altaffilmark{5}, Knox S. Long\altaffilmark{6}, 
John C. Raymond\altaffilmark{7}, and Stephen P. Reynolds\altaffilmark{4}}

\altaffiltext{1}{Department of Physics, Middlebury College, Middlebury, VT 05753; winkler@middlebury.edu}  
\altaffiltext{2}{NASA Goddard Space Flight Center, Greenbelt, MD 20771; brian.j.williams@nasa.gov}
\altaffiltext{3}{Department of Physics and Astronomy, Johns Hopkins University, 3400 N. Charles St., Baltimore, MD 21218; wpb@pha.jhu.edu}
\altaffiltext{4}{Physics Department, North Carolina State University, Raleigh, NC 27695; kborkow@ncsu.edu, reynolds@ncsu.edu}
\altaffiltext{5}{Department of Physics, Astronomy and Geosciences, Towson University, Towson, MD, 21252; pghavamian@towson.edu}
\altaffiltext{6}{Space Telescope Science Institute, 3700 San Martin Drive, Baltimore MD 21218; long@stsci.edu}
\altaffiltext{7}{Harvard-Smithsonian Center for Astrophysics, 60 Garden Street, Cambridge, MA 02138; jraymond@cfa.harvard.edu}

\begin{abstract}

We report results of infrared imaging and spectroscopic observations
of the SN\,1006 remnant, carried out with the \spitzer\ Space
Telescope. The 24\,$\mu$m image from MIPS clearly shows faint
filamentary emission along the northwest rim of the remnant shell,
nearly coincident with the Balmer filaments that delineate the
present position of the expanding shock. The 24\,$\mu$m emission traces the Balmer
filaments almost perfectly, but lies a few arcsec within, indicating an origin
in interstellar dust  heated by the shock.  Subsequent decline in the IR behind the shock is presumably due largely to grain destruction through sputtering.  The emission drops far more rapidly than current models predict, however, even for a higher proportion of small grains than would be found closer to the Galactic plane.  The rapid drop may result in part from a grain density that has always been lower---a relic effect from an earlier epoch when the shock was encountering a lower density---but higher grain destruction rates  still seem to be required.  Spectra from three positions along the NW filament
from the IRS instrument all show only a featureless continuum,
consistent with thermal emission from warm dust. 
The dust-to-gas mass
ratio in the pre-shock  interstellar medium is  lower than 
that expected for the Galactic ISM---as has also been observed in the analysis of IR emission from other SNRs but whose cause remains unclear.  
As with other SN\,Ia remnants, SN\,1006
shows no evidence for dust grain formation in the supernova ejecta.
\end{abstract}


\section{Introduction}
The supernova of 1006 C.E. is generally recognized as the brightest
stellar event in recorded human history, with surviving contemporary
reports from China, Japan, Korea, the Arab world, and in Europe as far
north as St.\ Gallen, Switzerland (latitude 47.4\degr\ N), all despite
the supernova's southern declination of Decl.\,$(1006) = -38.5 \degr$
\citep{stephenson10}.  Throughout much of the twentieth century,
SN\,1006 languished in relative obscurity compared with the attention
lavished on the other eleventh-century supernova, SN\,1054, that gave
rise to the Crab Nebula.  But once \citet{goldstein65} called
attention to the event and \citet{marsden65} constrained its location,
radio emission from the SN\,1006 remnant was soon discovered by
\citet{gardner65}.  It has since become a widely observed and
important object: X-rays from SN\,1006 were discovered by
\citet{winkler76}; optical emission by \citet{vandenbergh76}; and
(following a somewhat checkered history) TeV $\gamma$-rays were
definitively detected and mapped by the HESS collaboration
\citep{acero10}.  \citet{koyama95} discovered that the NE and SW limbs
of SN\,1006 have a featureless, power-law X-ray spectrum, providing
the first conclusive evidence that electrons can be accelerated to TeV
energies in supernova shocks and cementing the long-suspected
connection between supernovae and cosmic rays.  Yet no infrared
emission has previously been reported from SN\,1006.

The SN\,1006 remnant has a limb-brightened shell, 30\arcmin\ in
diameter with strong bi-lateral symmetry in radio and X-rays.  Its distance is well determined at
2.2 kpc \citep{winkler03}, and it is located high above the Galactic
plane ($b_{{\rm II}} = 14.6\degr,\ z=550\;$pc).  Its high-latitude
location leads to relatively low foreground absorption.  SN\,1006
itself was almost certainly a Type Ia event, based on its location,
environment far from any recent star formation, apparent absence of
any compact remnant, and implication from Chinese records that it
remained visible for several years \citep{stephenson02}.  The optical
emission from its remnant consists solely of Balmer-dominated
filaments \citep[e.g.,][]{schweizer78, kirshner87, ghavamian02}, which
indicate that it is expanding into an environment with relatively low
density that is at least partially neutral, another signature of SN\,Ia
remnants.

 Supernovae and their remnants are
critical to  dust  grain formation,
modification, and destruction.  Dust grains may condense from metal-rich
SN ejecta, and the fast shock waves in SNRs can lead to sputtering
through collisions with shock-heated gas that erode and eventually
destroy dust grains \citep[e.g.,][]{dwek80, draine03, sankrit10}.
Although these processes have been observed in a number of SNRs,
SN\,1006 nevertheless provides a unique perspective due to its
location high above the Galactic plane in a relatively pristine region
of the ISM.  There is growing observational evidence for the formation
of dust in the ejecta from core-collapse SNe, including SN\,1987A
\citep{wooden93, dekool98, matsuura11}, Cas A \citep{rho08, nozawa10}, and the Crab Nebula \citep{gomez12b}, and in
the late-time emission from a number of more distant SNe,
\citep[e.g.,][]{kotak09, szalai11, meikle11}.

Recent models by \citet{nozawa11} predict that small dust grains can also form in the ejecta from Type Ia SNe, and that $10^{-3} M_\sun$ of dust can survive for $\sim 1000$ years.  However, 
 no dust associated with ejecta has as yet been observed in any of historical remnants of the SN\,Ia: Tycho  (SN\,1572), Kepler SN\,1604), or RCW86 (probable SN\,185 remnant), 
all of which have been widely observed in the infrared.
Mid-IR emission from both Tycho and Kepler was detected from the
InfraRed Astronomy Satellite \citep[IRAS,][]{braun87}, and both have
been observed in detail from ISO \citep{douvion01}, {\em Spitzer}
\citep{blair07,williams12}, {\em Herschel} \citep{gomez12a}, and other
missions.  RCW86 was also detected from IRAS \citep{arendt89,
  saken92}, and \citet{williams11b} have recently presented detailed
observations with {\em Spitzer}.  \citet{borkowski06} have also reported IR
observations for four SN\,Ia remnants in the Large Magellanic Cloud.
All these cases show thermal emission from dust grains and provide
evidence for grain destruction in SN shocks, but none shows convincing
evidence for grain {\em formation} \citep[see especially][]{gomez12a}.

In this paper we report the first detection of IR emission from the
SN\,1006 remnant.  In Sections 2 and 3 we describe the results of an imaging
mapping campaign with the {\em Spitzer} Space Telescope, which reveals
a filament of 24\,$\mu$m emission along the NW limb of the shell, plus very faint, diffuse emission elsewhere.  In Section
4 we describe the spectroscopic observations from the InfraRed
Spectrometer (IRS) on {\em Spitzer}, and in Section 5 we present a
discussion in which we fit dust models to both sets of data and assess the total dust mass and dust-to-gas mass ratio.  Finally, we give a brief summary of our results in Section 6.

\section{MIPS Observations}

We mapped SN\,1006 twice with the Multiband Imaging Photometer for
{\em Spitzer} \citep[MIPS,][]{rieke04} as part of program 30673: on
2007 March 5 and 2007 August 22.  In order to cover the entire
30\arcmin\ shell and the surrounding background field, each
observation required a pair of overlapping scan maps, which were taken
consecutively.  The March AORs (18724608 and 18725120) were carried
out during a ``cold" campaign and included observations at 24$\mu$m,
70$\mu$m, and 160$\mu$m, while the August ones (18724864 and 18725376)
were during a ``warm" phase of the cryogenic mission and gave only
24$\mu$m and 70$\mu$m data.  Both pairs of scans consisted of 14 legs
of length 0.75\degr; offsets of 148\arcsec\ were used for the March
pair, while the August pair used offsets of 111\arcsec\ to give
greater exposure of SN\,1006 itself at a cost of somewhat less
surrounding background.  We began with Basic Calibrated Data (BCD)
files from the {\em Spitzer} Heritage Archive that had been processed
through version 18.12 of the {\em Spitzer} pipeline, and we then
performed further flat-fielding by taking a median of the hundreds of
frames from each AOR, and dividing all the frames from that AOR by the
normalized flat.  The estimated zodiacal light was then subtracted
from each frame, and the maps were mosaicked together onto a standard world
coordinate system, 50\arcmin\ square with 1\arcsec\ pixels, using the
MOPEX software provided by the {\em Spitzer} Science Center.  Finally,
we used DAOPHOT in IRAF\footnote{IRAF is distributed by the National
  Optical Astronomy Observatory, which is operated by the Association
  of Universities for Research in Astronomy, Inc., under cooperative
  agreement with the National Science Foundation.}  to remove the vast
majority of point sources (mainly stars) from the mosaic image, in
order to better assess the diffuse emission from SN\,1006 itself.

The full 24\,$\mu$m mosaic is shown in Fig.~1, with both the original
and star-subtracted versions in panels $a$ and $b$, respectively.  
These clearly show the first detection of SN\,1006 in the
mid-infrared.  For reference, we also show images of the identical
SN\,1006 field in \ha, X-rays, and radio in other panels of the
figure.  The most prominent IR feature is the thin filament along the
NW rim of the shell.  Even this brightest emission is quite faint,
with peak surface brightness levels above the background of
0.25\,-\,0.3 MJy sr$^{-1}$, only about 6\% of the background emission
and fainter than either the Tycho or Kepler SNRs by more than a factor
of 20.  
The integrated flux from this filament in
the 24$\mu$m band is $F(24\mu{\rm m}) = 130 \pm 30\;$mJy, where the
precision is limited by our tracing of the filament and ability to
assess the background accurately, since there is diffuse background as
well as faint residual stars in the ``star-subtracted" mosaic.

We used a similar approach for the 70 $\mu$m data, except that here
the mosaic has a scale of $4\arcsec\perpix$.  Despite our
flat-fielding all the data using median frames for each AOR, artifacts
along the scan directions are nevertheless evident in the final
mosaic, shown in the bottom left panel of Fig.~1.  There is the
slightest hint of an enhancement in the region of the NW filament, but
this does not trace the narrow filament and is barely larger than
background fluctuations in the vicinity.  We give only an upper limit
for the average surface brightness in the region of the filament:
$S(70\mu{\rm m}) < 0.85$ MJy sr$^{-1}\ (3\sigma)$, and an integrated
flux $F(70\mu{\rm m}) < 800$ mJy.  The 160 $\mu$m images, compiled
using only the two scans from 2007 March for which 160 $\mu$m are
available, are quite noisy and show no discernible emission from SN\,1006.

\section{Imaging Analysis and Multi-wavelength Comparisons}
\subsection{The Northwest Filament}
The NW filament corresponds closely with the optical filament that
appears only in the hydrogen Balmer lines \citep{vandenbergh76,
  winkler03, raymond07}, as shown in the deep continuum-subtracted
image shown in the lower right panel of Fig.~1.  Such ``nonradiative"
filaments characterize fast shocks expanding into a low-density,
partially neutral environment, and that are generally associated with
relatively young remnants from Type Ia supernovae such as SN\,1006.
Balmer emission from these filaments has two components, both the
result of neutral H atoms that cruise through the shock.  In the hot
post-shock environment, the neutrals can be excited and decay to
produce narrow lines (with a width characteristic of the pre-shock
temperature), or they can undergo charge exchange with hot protons to
produce broad lines whose width is closely related to the post-shock
proton temperature \citep[e.g.,][] {chevalier80, ghavamian02, heng10}.
Since the lifetime of neutral atoms in the hot post-shock environment
is very short, the Balmer filaments can occur only {\em immediately}
behind the shock, and thus delineate the current position of the
shock.

To investigate the post-shock characteristics of emission along the NW
filament in various bands, we have compared 24$\mu$m, \ha, and X-ray
images.  We have used a higher resolution \ha\ image, taken from the 0.9\,m telescope
at CTIO on 2002 March 21, for which the stars were subtracted using a matched
continuum image \citep[for details, see][]{winkler03}, instead of the one shown in Fig.~1c.
Likewise, in X-rays we 
have used the 0.5-2.0 keV data from a 90 ks observation taken 2001
April 26-27 with the {\it Chandra} ACIS-S \citep[for details,
  see][]{long03}, rather than the mosaic of shorter ACIS-I exposures used in Fig.~1e.  We first checked the alignment of all three
images\footnote{For checking the \ha\ image we of course used the
  version {\em before} continuum subtraction; the coordinates for the
  subtracted image are identical.}  using four near point sources
common to all: one bright star and three distant galaxies, and found
their positions in all three bands to agree within 0.5\arcsec.  To
simulate images in all three bands at a common epoch, we used the $\mu
= 280 \pm 8 \masy$ proper motion rate for the NW Balmer filaments,
measured relative to the center of curvature at ${\rm R.A.\, (2000.) =
  15^{\rm h} \, 03^{\rm m}\, 16\fs2,\ Decl.\,(2000.)}  = -42\degr \,
00\arcmin \, 30\arcsec $ by \citet{winkler03}, to extrapolate the
\ha\ and X-ray images from their original epochs (2002.2 and 2001.3,
respectively) to the 2007.5 mean epoch for the {\em Spitzer} MIPS
scans.  \citep[Recent measurements from {\it Chandra} find X-ray
  proper motions along most of the NW rim of SN\,1006 that agree with
  those measured for the optical filaments,][]{katsuda12}.

In Fig. \ref{opt_ir_xray} we show a three-color combination that
compares the 24\,$\mu$m (red), \ha (green), and X-ray (blue) images of
the NW filament, and in Fig. \ref{profiles} we plot radial profiles
taken perpendicular to the filament, integrated over the four
azimuthal regions outlined in Fig. \ref{opt_ir_xray}.  It is clear
that the 24\,$\mu$m emission has a broader spatial profile that peaks
3 - 5 arcsec behind the Balmer filament, as best illustrated in region
D of Fig. \ref{profiles} where the profiles are the cleanest.  The
filament is also noticeably broader in the IR than in \ha, with FWHM
$\gtrsim 15\arcsec$.  This is not simply a resolution effect.  The
 resolution of the 24\,$\mu$m mosaic image---measured from profiles of several isolated stars---is 6\arcsec\ 
(FWHM), consistent with the diffraction limit, but  small
compared with the width of the filament.

We attribute this IR emission to dust from the ISM that enters the
shock, where it is heated through collisions with hot post-shock gas,
produces IR emission, and is then destroyed.  Somewhat further behind
the shock, soft, primarily thermal, X-ray emission appears, rising to
its peak about 10\arcsec\ behind the shock, just as the 24\,$\mu$m
emission is decaying away.  This indicates the longer time scale for
producing the highly ionized species that radiate efficiently and
dominate the X-ray emission \citep{long03}.  We can estimate the time
scales through simple kinematics: the primary shock is moving outward
to the NW at $\mu = 280\pm 8 \masy$, and if we assume the nominal
compression ratio of 4 between the post-shock and pre-shock densities,
the post-shock material is dropping behind the shock at speed $v_s/4$, or $\mu = 70 \masy$ (relative to the shock). 
Therefore, the IR peak occurs
some 60 yr after that material was shocked, and the X-ray peak occurs
$\sim 150$ yr after the shock.  We shall discuss the emission profiles further in Sec. 5.1 of this paper.

In 2012 March, all-sky survey data from the Wide-field Infrared Survey
Explorer \citep[WISE,][]{wright10} were released.\footnote{ see
  http://wise2.ipac.caltech.edu/docs/release/allsky/.}  Examination of
data from the SN\,1006 region in all the WISE bands (3.4, 4.6, 12 and
22 $\mu$m) shows emission from SN\,1006 only in the 22\,$\mu$m band.
 Not surprisingly, the WISE 22\,$\mu$m image is
very similar to the {\em Spitzer} 24\,$\mu$m one, and clearly shows
the NW filament.  However, the WISE data have somewhat lower angular
resolution and signal-to-noise than that from {\em Spitzer}.   
The non-detection of emission in the shorter-wavelength {\it WISE} bands is 
consistent with the lack of short-wavelength
emission in other Type Ia remnants \citep{borkowski06}.
Close examination of the images from the Improved Reprocessing of the IRAS
Survey \citep[IRIS;][]{miville05} also shows traces of the NW filament
in the co-added 25\,$\mu$m image; however, it is unlikely that this
faint feature would ever have been independently discovered in these
low-resolution (1.5\arcmin) maps.   

\subsection{Other Faint IR Emission}
In addition to the narrow filaments along the NW rim of the shell,
there is patchy, diffuse emission within much of the southern part of
the SN\,1006 shell, clearly visible in the 24\,$\mu$m mosaic and less
so in the 70\,$\mu$m one.  The morphology of this emission is similar
at both 24\,$\mu$m and 70\,$\mu$m, and is also similar to patchy
emission extending well
outside the SN\,1006 shell, especially to the south and east.   In the southern portion of SN\,1006,  there is also faint, diffuse emission in both \ha\ and X-rays  (Fig.~1c and e), but unlike along the NW rim, the morphology in the south is only  vaguely similar to that seen in the IR\@.  In the absence of further data, we cannot determine whether the faint IR emission is associated with SN\,1006 itself or stems from  cool foreground dust.


In X-rays, SN\,1006 shows knots of emission interior to the shock that are thought to arise largely from SN ejecta,  
but there are no obvious small-scale IR features that correlate with these X-ray-emitting ejecta knots.  
 Hence we find no evidence for the formation of dust in SN\,1006 ejecta.  This is somewhat surprising in view of theoretical predictions for dust formation in SN\,Ia ejecta \citep{nozawa11}, but is consistent with observations of other Type Ia SNRs \citep{borkowski06,williams12}.

\section{Spectroscopic Observations}

We conducted IRS observations, all in Staring mode for about 90 min each,  at three locations
along the NW filament. We show the slit positions of these  IRS pointings, labeled 1-3, in
Fig.~\ref{irs_overlays}, overlaid on the MIPS 24 $\micron$ image of
the filament. As noted earlier, the filament is quite faint in the IR (its average
surface brightness is only about 6\% above the background at 24
$\micron$), and it was undetected in the SL (5.5-14 $\micron$) spectrum.
However, faint emission is detected in the LL order 2 (14-19.3
$\micron$) spectrum, where the dust continuum begins around 17
$\micron$ and continues to rise throughout the LL order 1 (19-40
$\micron$) spectrum.  We see no evidence for line emission anywhere in
the spectrum, consistent with the non-radiative nature of the blast
wave shock at this location \citep{dwek87, dwek96, williams11a}.

The data used in our analysis (AOR IDs 28138240 and 24142080) were
processed with \spitzer\, calibration pipeline version S18.7.0.  Using
the IDL CUBISM software \citep{smith07}, we assembled the 72 BCD files
from positions 1 and 2 along the NW filament into two data cubes, one
for LL1 and one for LL2.  The IRS spectrum from position 3 is similar, but we have omitted it from our subsequent analysis  because it falls on a
region where different abundances were required to fit the X-ray spectrum \citep{long03}.  Hot pixels
affect the BCDs (especially beyond 35 $\micron$), so we median
smoothed the BCDs before creating the data cubes using the
FILTER\_IMAGE routine from the IDL Astronomy User's library.  The
smoothing replaces the value of each pixel with the median of the
surrounding 3 pixels in a moving box and significantly mitigates the
hot pixels in the resulting data cubes.  While the smoothing reduces
the spatial resolution of the data somewhat, the emission from the
filament is highly uniform and we expect minimal loss of information.

After combining the BCDs within CUBISM, we extracted LL1 and LL2
spectra from the NW filament using positions 1 and 2 marked in
Fig~\ref{irsoverlays}.  These two positions are only $\sim
30\arcsec$ apart and their IRS spectra appear virtually identical;
therefore, we have combined them into a single object spectrum.  To
estimate the sky background, we extracted separate LL1 and LL2 spectra
from locations just outside SN\,1006 (marked in Fig~\ref{irsoverlays}).
We averaged the two sky spectra in each order and combined them to
produce a final sky spectrum.  We subtracted this spectrum from that
of the SN\,1006 filament to obtain the final spectrum shown in
Fig.~\ref{irsfit}.  We describe the fits to this spectrum in the
following section.

\section{Discussion}

\subsection{Model Fits to the Data}

Infrared emission in typical SNR shocks is produced entirely by
collisional heating of dust grains (Dwek \& Arendt 1992), where
energetic electrons and ions collide with the grains, heating them to
temperatures of 50-100 K\@.  Collisions with protons and heavier ions are also responsible for grain  sputtering, 
reducing the size of large grains and destroying many small grains
entirely. We have developed models for such dust that
self-consistently calculate both the heating and the destruction of
grains in the post-shock gas as a function of the density of the hot
plasma, $n_p$, the electron and ion temperatures, and the shock
sputtering age $\tau = \int_0^t {{n_p}} dt$ \citep{borkowski06}.  For
an arbitrary input grain size distribution, we can use these models to
calculate the IR spectrum and the post-shock profile that results as
grains are heated, and eventually destroyed.

The  electron temperature,  $T_e$, and the
ionization timescale can be determined from a non-equilibrium
ionization analysis of the X-ray spectra, as \citet{long03} did for
the NW limb using {\it Chandra} data.  We have revisited that
analysis, taking into account hydrocarbon contamination of the ACIS
detectors that was unrecognized at the time of the original analysis,
and using more reliable atomic data.  We first redetermined the
absorption column density $N_{\rm H}$ from the reprocessed
synchrotron-dominated eastern limb data \citep[region NE-1 of][]
{long03}, fit using XSPEC \citep{arnaud96}  with a pure synchrotron {\sl srcut} model to
obtain $N_{\rm H}$ = (8.2 $\pm$ 0.8) $\times$ 10$^{20}$ cm$^{-2}$.  We
fixed $N_{\rm H}$ at this value in fitting Long et al's region NW-1,
which includes our IRS regions 1 and 2. We obtained reasonable fits
using a solar-abundance plane-shock model ({\sl vnpshock}), and
obtained values of $kT_e = 0.9$ keV, ionization timescale $n_et = 5.3 \times
10^9\,{\rm cm^{-3}\,s} = 170\,{\rm cm^{-3}\,yr}$, and emission measure $n_e M_g = 6.2 \times
10^{-3}\ M_\odot$ cm$^{-3}$.  This is a somewhat higher electron
temperature than the $kT_e = 0.6$ keV found by \citet{long03}, but is still far below the
 post-shock proton temperature of $kT_p = 16$ keV, inferred
from the Balmer line profiles by \citet{ghavamian02}.  \citep[][also measured a similar value for $T_e$ at a different position along the NW rim.]{vink03}.  We then applied a
Coulomb-heating model for collisions between electrons and protons
behind the shock, and found that over the relatively small region
we consider here, the mean proton and electron temperatures are 16 keV
and 1.0 keV, respectively. We use these temperatures for the
remainder of this work.

Our models also require as input a distribution of grain sizes in the
ISM ahead of the shock.  A commonly used distribution for the Galaxy
is that of \cite{weingartner01}.  Their models are appropriate
for the diffuse ISM in the disk of the Galaxy, where the
total-to-selective extinction ratio $R_V \equiv A_V/E(B-V) = 3.1$, and
for dense clouds where $R_V > 3.1$.  At the location of SN\,1006 some
500 pc above the Galactic plane, however, the abundance of dust---both in total
amount and in the distribution of grain sizes---is likely very
different from that near the  Galactic plane. 
There is significant evidence for much lower values for $R_V$ at high Galactic latitudes \citep{larson05}.  
\citet{mazzei11} have calculated
grain size distributions using the same formalism as
\citet{weingartner01} for over 70 lines of sight to stars with
``anomalous extinction sightlines.'' Particularly relevant for our
work, they extend these models down to low-density environments with
$R_{V} = 2.0$. We use the model from Table 3 of \citet{mazzei11},
with $R_{V} = 2.0$ and $10^{5} b_{c} = 5.0$ (this is a
parameterization in their models that corresponds to the number of
small carbonaceous grains). The grain size distributions corresponding
to lower values of $R_V$ are steeper than their ``typical'' Milky Way
counterparts, i.e., they have an excess of small grains relative to
large ones, resulting in a larger fraction of  the total mass in
grains being destroyed via sputtering for a given shock, as well as a
different gas density required to produce the observed spectrum.

In our models, the spectrum from warm dust is most sensitive to the gas density in post-shock plasma, 
so we have used proton density $n_p$ as the principal
free parameter.  Using the grain size distribution described above, we
have fit the IRS spectrum over the range 21 - 37 $\mu$m and obtain a best fit for 
a density of $n_{p} = 1.4 \pm 0.5 \, {\rm cm}^{-3}$, shown in blue in Fig.~\ref{irsfit}.  The 21 - 37 $\mu$m range comprises all the  data for which the signal-to-noise ratio is high enough to be useful.  The uncertainty quoted is a formal 90\%-confidence limit based on statistical uncertainties only, determined from 
 the rms deviations from an arbitrary smooth polynomial fit to the spectrum in this range.   We have made no provision for   systematic errors in the IRS spectra, which are difficult to quantify.  
 
There is good reason to believe that the actual density is near the lower limit of the fitted range, near  $n_p =1 \; {\rm cm}^{-3}$.  
From an analysis of the thermal X-ray emission along the NW rim based on {\em Chandra} observations, \citet{long03} found a post-shock electron density $n_e \sim 1\, {\rm cm}^{-3}$.  Analysis of X-ray data from {\em XMM-Newton} using a  different modeling approach led \citet{acero07} to a {\em pre}-shock density value of $n_{0} \sim 0.15 \, {\rm cm}^{-3}$, equivalent to a post-shock density of $n_p \sim  0.6 \, {\rm cm}^{-3}$ for a standard compression ration of 4.  
And finally, an estimate based on
the thickness of the H$\alpha$ filaments in an image from the {\it
  Hubble} Space Telescope by \citet{raymond07}, and subsequently
corrected by \citet{heng07}, gave $n_{0} = 0.15-0.3$, and thus a post-shock density $n_p \sim 1\, {\rm cm}^{-3}$. 
In 
Fig.~\ref{irsfit}  we also show, in red, a model with $n_p = 1.0\, {\rm cm}^{-3}$, which gives an acceptable fit to the IRS data.   Finally, we note  that the predicted 70 $\mu$m surface brightness for any of our models is
$\lesssim 0.25\, {\rm MJy\; sr^{-1}}$, well below the upper limit
reported in Section 2.

We have also calculated the spatial profile for the decay of IR emission as
grains are destroyed through sputtering in the hot, post-shock
environment, assuming that the shock has been advancing through an ambient medium of constant density.  In comparing the model with the observed radial profiles, we
find that the observed decay is far more rapid than the sputtering model
predicts, as illustrated in Fig.~\ref{profile_models}.   Even if we
assume a grain-size distribution heavily weighted toward small grains to
give  $R_{V} = 2.0$, for a density of $n_{\rm  H} = 1\, {\rm cm}^{-3}$
the sputtering rate would need to be artificially  enhanced  by a factor 6 to 8 to
match the observed profiles. 

However, it is almost certain that the shock along the NW rim of SN\,1006 has {\em not} 
been passing through a uniform-density environment.  The 
ambient density to the NW is clearly higher than that around
the remainder of the shell, as evidenced by the following facts:  (1) the flattening of the shell
to the NW; (2)  the fact that the proper motion along most of the NW limb is
only about 60\% as high as that to the NE \citep{katsuda12}; (3)   the
density contrast around the shell required from X-ray observations  \citep{long03, acero07}; and (4) the mere fact that there is significant IR emission along the NW limb, yet little or none elsewhere.  
Since the shell radius to the
NW is only about 15\% smaller than elsewhere, the shock must have
encountered the denser region there relatively recently.  As a result,
the post-shock density profile, for both gas and dust, should drop  as
one moves inward from the shock.   Therefore the decrease in IR emission
behind the shock is likely due   in large part to the fact that farther inward the
grain density has always been lower, representing an earlier epoch in
the evolution of SN\,1006.   We suggest that this may account for the
rapid drop in IR emission $\sim 20\arcsec - 35\arcsec$ behind the shock
(Fig.~\ref{profile_models}).   For a density  $n_p = 1\,{\rm cm}^{-3}$ (equivalent to $n_e = 1.2\,{\rm cm}^{-3}$), the shock age from X-ray spectroscopy is about $170/n_e = 140$ years, suggesting that the shock was encountering considerably lower densities earlier.  
Detailed modeling of such a
scenario is beyond the scope of the present paper. 

Nevertheless, the fact that the observed post-shock drop in IR emission is so much sharper than predicted by the models, even close to the shock, probably indicates that current models significantly underestimate  the actual dust sputtering rate.   Effects that could be responsible include an even greater proportion of small grains, grains of high porosity, or a grain composition that leads to higher sputtering rates.  Enhanced sputtering rates are not without precedent; \citet{serra-diaz-cano08} found that hydrogenated dust grains (perhaps the most common type of grains in the ISM, at least for carbonaceous materials) are sputtered at rates several times higher than their pure amorphous counterparts.  The spectral data are insufficient to determine whether the carbonaceous grains in SN\,1006 are hydrogenated or amorphous.   

\subsection{Dust Mass of the NW Filament}

Almost independent of the models, we can determine the 
 total amount of dust radiating in the NW filament, provided that the integrated IR  luminosity is relatively  well-known.  We have used the measured IR flux, $F(24\mu{\rm m}) = 130 \pm 30\;$mJy (Section 2), and assumed that the overall shape of the
dust spectrum observed with IRS (shown in Fig.~\ref{irsfit}) is applicable along
the entire length of the filament to obtain  $L_{\rm IR} = (1.7 \pm 0.9)  \times 10^{34} \,d_{2.2}^{ \:2}\; {\rm ergs\;s^{-1}}$, where we have scaled to a distance of 2.2 kpc: $d_{2.2} \equiv d/2.2\;{\rm kpc}$.    Our models assume
optical constants from \citet{draine84} for graphite and amorphous
silicate grains in the ISM, with relative proportions and grain-size
distributions from \citet{mazzei11}.    As in \citet{williams08}, we then calculate  the dust mass
in the NW filament as $M_{\rm dust} = (1.1 \pm 0.6) \times {10^{ - 5}}\, d_{2.2}^{  \: 2}\; \msun$, where the   
uncertainty limits for both $L_{\rm IR}$ and $M_{\rm dust}$ include the
overall 15\% uncertainty in the {\it Spitzer} MIPS extended source
flux calibration.  Of course, this method of determining the total dust
mass is insensitive to any cold dust---undetectable with {\it
  Spitzer}---but it is difficult to imagine how a significant fraction
of interstellar dust (consisting almost entirely of small grains, $a<0.5\,\mu$m) could remain cold within the hot plasma behind such a fast
shock. 
Our value for $M_{\rm dust}$ includes only the dust now radiating along the NW filament; 
 the overall faintness 
of the rest of the SNR at 24\,$\mu$m, combined with the lack of detection at 70
$\mu$m, precludes  determination of a dust mass for the entire remnant.

\subsection{The Ambient Dust-to-Gas Mass Ratio}

In previous works \citep{borkowski06,   sankrit10, williams11a, williams11b}
we have used the density determination from IR data,
combined with the X-ray emission measure, to obtain a measure of the
amount of shocked gas present within a given emitting volume. We have
obtained the dust mass from the overall normalization to the dust
spectrum, so dividing the two gives the dust-to-gas mass ratio.  In
the case of SN\,1006, however, the IRS spectra do not strongly
constrain the density. Thus, we employ an alternative approach.

\citet{laming96} examined the ultraviolet spectrum of a section of the NW
filament obtained with the {\it Hopkins Ultraviolet Telescope} ({\it
  HUT}) and measured 0.02 He\,II (1640 \AA) photons cm$^{-2}$
s$^{-1}$.   More recent work by Chayer (private communication)  has revised the
reddening towards SN\,1006 to be $E(B-V) = 0.09$, lowering the He\,II
flux to 0.017 photons cm$^{-2}$ s$^{-1}$\@. According to the Laming et al.\ models,
0.0067 He\,II photons come out for each H atom crossing the shock. For
the 200\arcsec\ section of the filament observed (corresponding to a
distance of $6.5 \times 10^{18}$ cm at $d=2.2$ kpc), the He II flux
measured at Earth would be
\[{F_{{\rm{He\, II}}}} = \left( {0.0067\;\frac{{{\rm{photons}}}}{{{\rm{H\; atom}}}}} \right)\left( {6.5 \times {{10}^{18}}\,{\rm{cm}}} \right)\frac{{{n_0}{\upsilon _s}l}}{{4\pi {d^2}}}\, ,\]
where $n_{0}$ is the pre-shock H number density, $v_{s}$ is the shock
speed, and $l$ is the line-of-sight depth through the emitting
material, all of which are presumed (for this purpose) to be
unknown. Taking $F_{{\rm He\, II}} = 0.017$ photons cm$^{-2}$ s$^{-1}$,
we can rearrange the above equation to give $n_{0}v_{s}l = 2.26 \times
10^{26}$ cm$^{-1}$ s$^{-1}$. The length of the NW-1 region is
$3\arcmin$, or $5.9 \times 10^{18}$ cm at $d=2.2$ kpc, so $1.3 \times
10^{45}$ H atoms per second enter the shock. If we assume a standard
post-shock flow of $70 \masy$, based on the proper motion measurements
of \citet{winkler03}, then the width of the IR filament corresponds to
a timescale of 150 years. During this time, the shock will have swept
up 6.3 $\times 10^{54}$ H atoms, or 7.6 $\times 10^{-3} \msun$ of gas. 
The uncertainties in the {\it HUT} measurement of the He II flux and in the
reddening correction are about 15\% each, while the uncertainty in
proper motion is only 3\% and the uncertainty in the relative atomic
rates is probably about 30\%, so we estimate an uncertainty in the gas
mass of $\sim 40\%$\@.  Using
the dust mass determined above, this leads to a dust-to-gas mass ratio
of $\sim (1.5 \pm 1.0) \times 10^{-3}$. The expected dust-to-gas mass ratio for
the Galaxy within the model of \citet{mazzei11} is 5.3 $\times
10^{-3}$, a factor of $\sim 3$ higher than what we measure. 

The dust-to-gas mass ratio in the ambient medium surrounding SN 1006
appears to be lower than that expected from dust models for the Galaxy,
including those at high Galactic latitudes. This has been the case for
virtually all SNRs studied, both in the Galaxy
\citep[e.g.,][]{blair07,arendt10,lee09} and in the Magellanic Clouds
\citep{borkowski06,williams06,williams11a}. The ultimate reason for this
remains a mystery, but in the case of SN\,1006 specifically, its location
within the Galaxy may provide a clue. At some 500 pc above the Galactic plane, SN\,1006
is located at the interface between the disk and halo \citep[see][for a review of gaseous galaxy halos and their interfaces to
galactic disks]{putman12}.  Accumulating evidence points to a lower dust content
there than in the Galactic disk \citep{ben-bekhti12}, coupled with a higher proportion of smaller
grains \citep{planck11}. An expected increase in the
sputtering rate for smaller grains might better match the rapid drop-off
in the 24\,$\mu$m profile behind the shock (Fig.~\ref{profile_models}) than the current dust
model, and it would also raise our inferred value of the pre-shock
dust-to-gas mass ratio, bringing it even more in line with the \citet{mazzei11}
value. We further note that any  dust that is too cold to be observed with {\em Spitzer} would increase the dust-to-gas mass ratio above our inferred value.

Our current determination of the dust-to-gas mass ratio is quite
uncertain, as we do not know the grain size distribution in the
pre-shock gas. It is possible that models with a larger dust-to-gas ratio
and a larger overabundance of small grains can be as viable as our
current model with $R_V=2.0$. This model degeneracy might be broken by
future joint IR, X-ray, and UV studies focused on accurate
determination of abundances of refractory elements in the postshock
gas.

\section{Summary}

{\it Spitzer} observations of the entire region surrounding the
remnant of SN\,1006 show a clear detection of the NW filament at 24
$\mu$m, the same filament that is prominent in H$\alpha$ emission. IRS
spectroscopy of this filament shows that the IR  is produced entirely by
continuum emission from warm dust grains, with a spectrum peaking
somewhere beyond 35 $\mu$m. The filament is not detected at 70 $\mu$m,
likely due to the significantly higher background IR emission.
The IR emission from
the NW filament reaches its peak just inside of the H$\alpha$ emission, while the
X-ray emission peaks still a little further behind. This is as
expected, since H$\alpha$ emission should trace the exact location
of the forward shock, while dust grains require a short amount of time
to be heated to temperatures observable by {\it Spitzer} and then are sputtered away, while the
post-shock gas requires even longer to become ionized enough to emit strongly
at {\it Chandra} energies.

Fits to the  {\it Spitzer} IRS spectrum taken at two positions along the NW filament, using
models of collisionally-heated dust grains, do not tightly constrain the post-shock density $n_p$.  The nominal best fit gives $n_{p} = 1.4 \pm 0.5\; {\rm cm}^{-3}$ (90\%-confidence); we have adopted  a model with $n_{p} = 1\; {\rm cm}^{-3}$, the value suggested from observations in other bands, for our subsequent analysis.   
The decay of the IR spatial profile behind the shock is far more rapid than expected from standard dust models.  This is may be due in part to a high proportion of small dust grains in the ambient ISM at the location of SN\,1006, high above the Galactic plane, and probably also results in part as a relic of the shock's having encountered a higher density region to the NW relatively recently.   Nevertheless, it appears that sputtering rates significantly higher than those predicted by standard models will be required to fully explain the rapid decline in IR emission behind the shock.

Calculations of  the dust-to-gas mass ratio in the ambient ISM indicate a value of approximately $1.5 \times 10^{-3}$, compared to the expected
value of 5 $\times
10^{-3}$ for the Galaxy at the latitude of SN\,1006. This difference of a factor of several is a phenomenon seen
in virtually all IR/X-ray analyses of SNRs. The remnant of SN\,1006
provides a rare opportunity to study the ISM high above the Galactic
plane, and further IR, X-ray, and UV studies are necessary to
understand this remnant.

\acknowledgments

This research is based on observations made with the \spitzer\ Space
Telescope, which is operated by the Jet Propulsion Laboratory,
California Institute of Technology under a contract with NASA.   We are grateful to the support staff at IPAC for their guidance with some of the subtleties of {\em Spitzer} data analysis.  We also acknowledge thoughtful comments from the anonymous referee, which have prompted us to, we hope, clarify much of this paper.  
Primary financial support for this project has been provided by NASA through RSA 1330031.  PFW
acknowledges  additional support from the NSF through grant
AST-0908566, and KJB acknowledges  additional support from NASA through grant NNX11AB14G.
\clearpage

\bibliographystyle{apj}




\clearpage 

\begin{figure}
\plotone{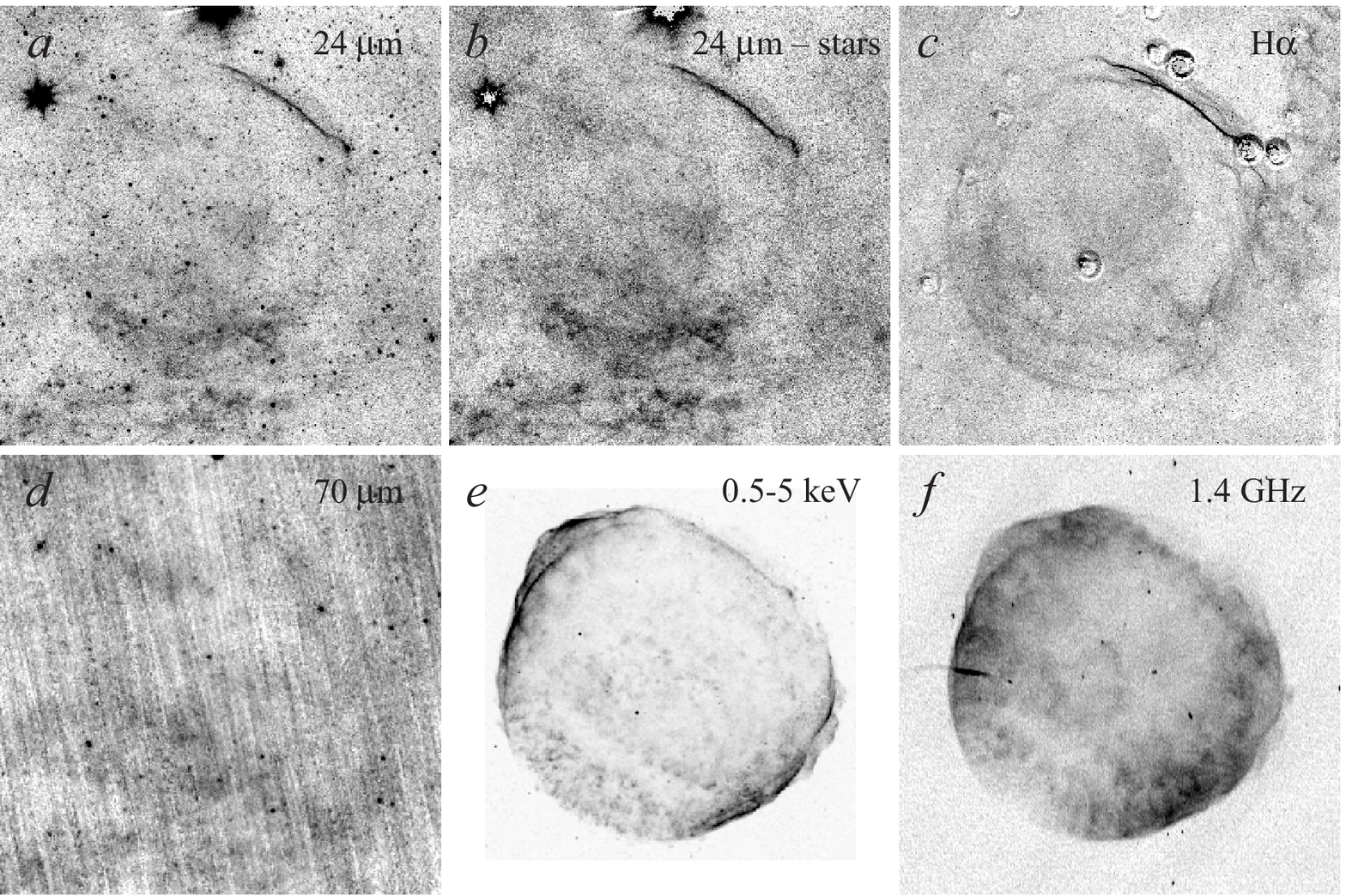}
\caption{Images of SN\,1006 in various bands. {\em Panel a:} Mosaic
  image at 24\,$\mu$m, assembled from all four MIPS AORs.  {\em b:} The same 24\,$\mu$m mosaic, after subtraction of most
  point sources, displayed with a harder stretch.  {\em c:}
  Deep \ha\ image, after continuum subtraction to reveal the extremely
  faint shell of emission to the S and E \citep[the NW filament is
    saturated in this display; the irregular donut-like patterns are
    artifacts, the residual images from bright
    stars---from][]{winkler03}.  {\em d:} Mosaic image at
  70\,$\mu$m, from all four MIPS AORs. {\em e:} {\em
    Chandra} ACIS-I mosaic of SN\,1006, from \citet{cassam-chenai08}; this
  image emphasizes the synchrotron-dominated emission in the NE and
  SW.  {\em f:} 1.4 GHz radio image, from \citet{dyer09}.
  All the images are aligned, with a field 40\arcmin\ square and
  oriented north up, east left. }
\label{mosaic}
\end{figure}

\begin{figure}
\plotone{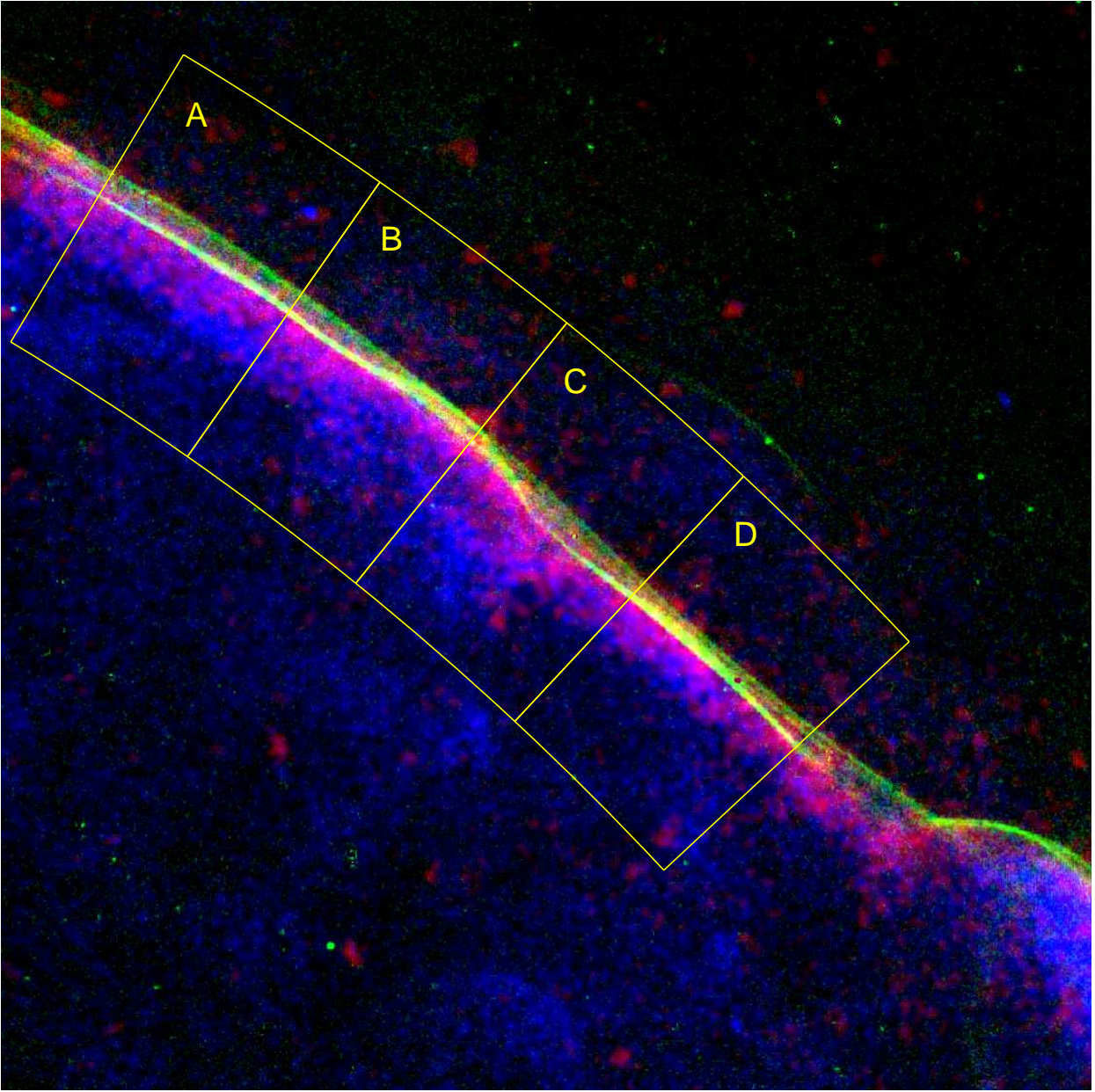}
\caption{This 3-color image of the NW portion of SN\,1006 shows the
  evolution in conditions behind the shock, which is propagating
  toward the NW (upper right corner).  \ha\ filaments 
  \citep[green, from][]{winkler03} form immediately behind the shock; 24\,$\mu$m
  emission (red) from warm dust trails slightly behind as the dust
  grains are first heated and then destroyed; X-rays  \citep[blue, 0.5 - 2.0
  keV, from][]{long03} peak further behind due to the ionization
  time to reach the species that are most effective at producing X-ray
  emission.  The field is 6.5\arcmin\ square; yellow regions indicate
  locations for the profiles shown in Fig.\ 3.}
\label{opt_ir_xray}
\end{figure}

\begin{figure}
\plotone{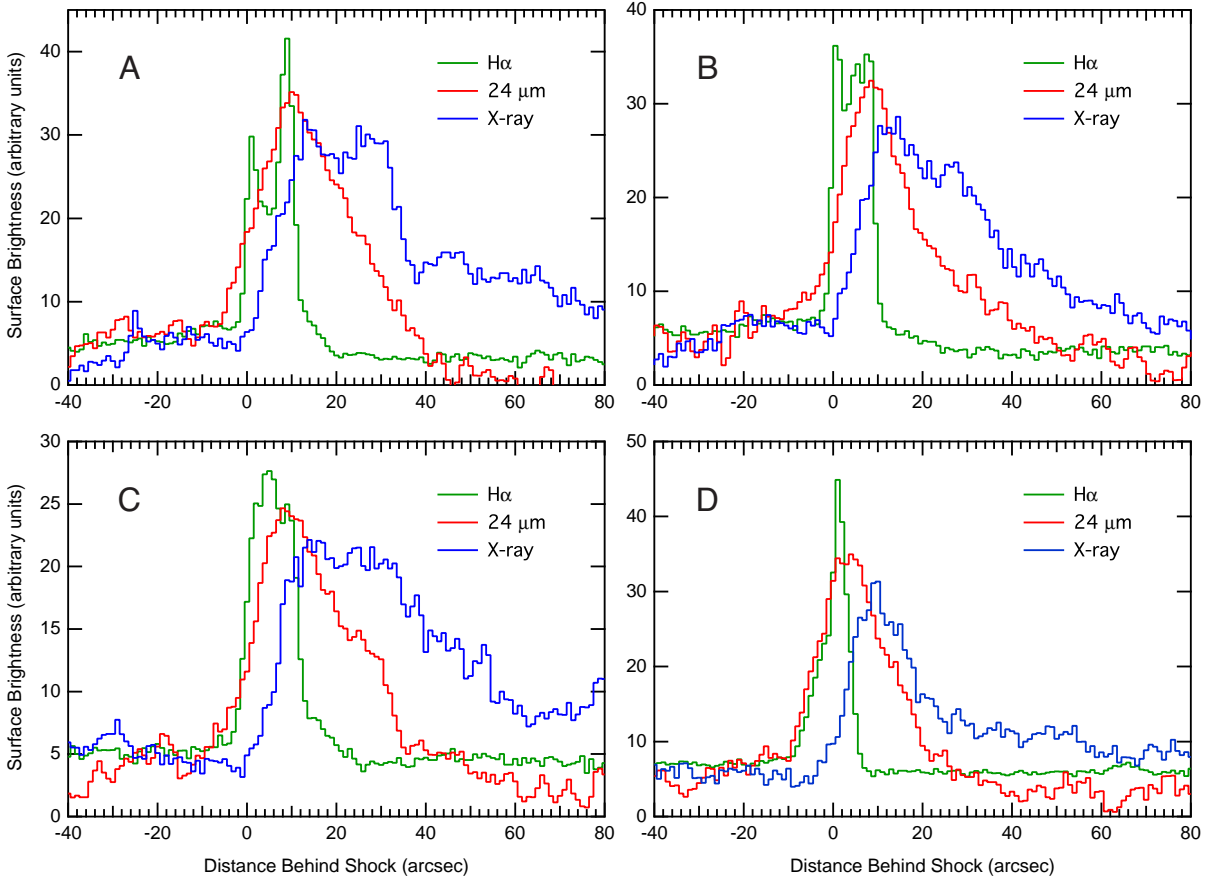}
\caption{Radial profiles of emission in \ha, 24\,$\mu$m, and X-rays, taken
  perpendicular to the shock, for the four regions shown in
  Figure~\ref{opt_ir_xray}.  (The data sources are the same as for that figure.)  In each case, the profiles run from outside the shell inward, and the shock location is arbitrarily taken as the point at which \ha\ emission first rises sharply.}
\label{profiles}
\end{figure}

\begin{figure}
\plotone{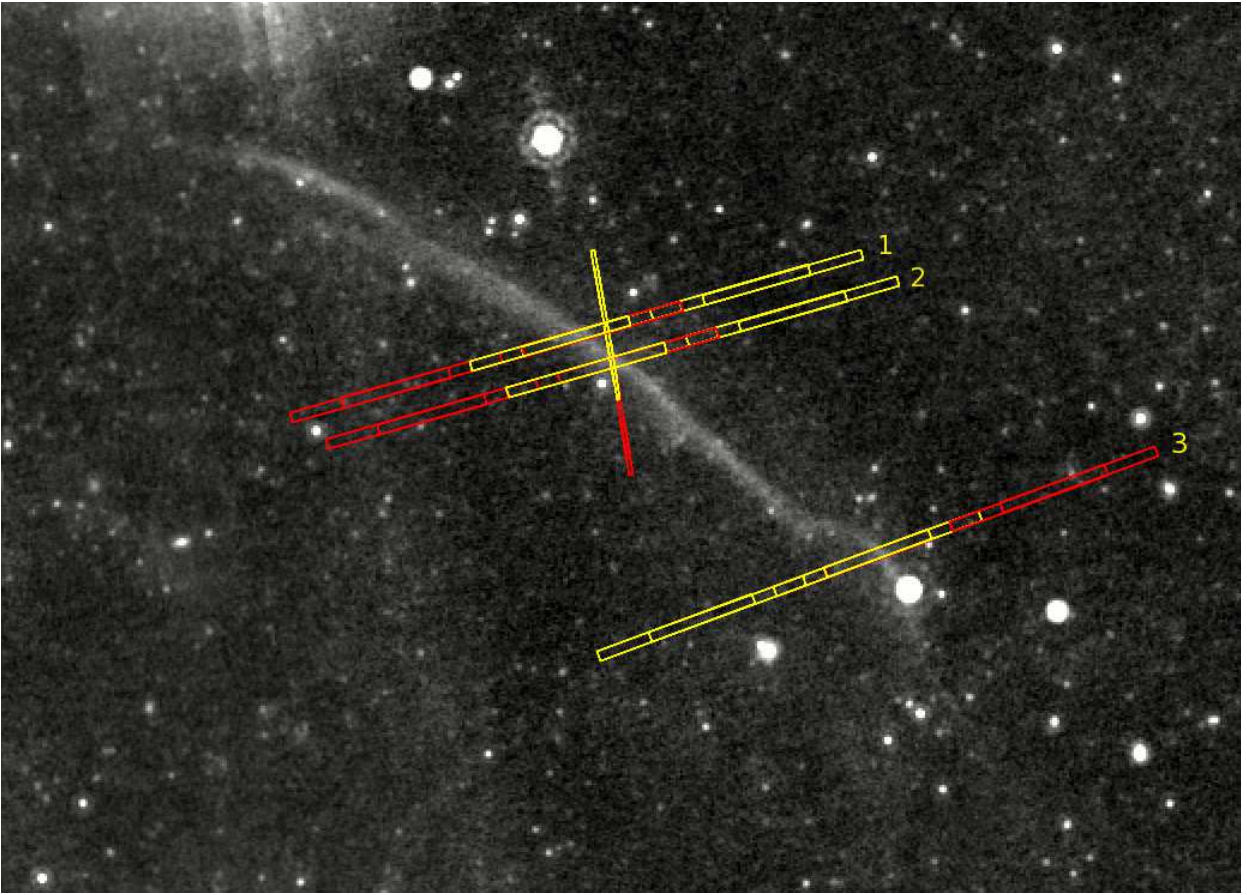}
\caption{\label{irs_overlays}IRS positions superposed on the 24\,$\mu$m
  image of the NW filament of SN\,1006. Red rectangles correspond to
  the second order (longer wavelength) spectrographs, while yellow
  corresponds to first order. The three numbered slit positions are from long pointings 
  with the IRS Long-Low instrument (14-40 $\mu$m). As we discuss in
  the text, we use positions 1 and 2 for our spectral modeling. The
  shorter, nearly vertical slit position is from the IRS Short-Low
  instrument, where no emission was detected.}
\label{irsoverlays}
\end{figure}

\begin{figure}
\plotone{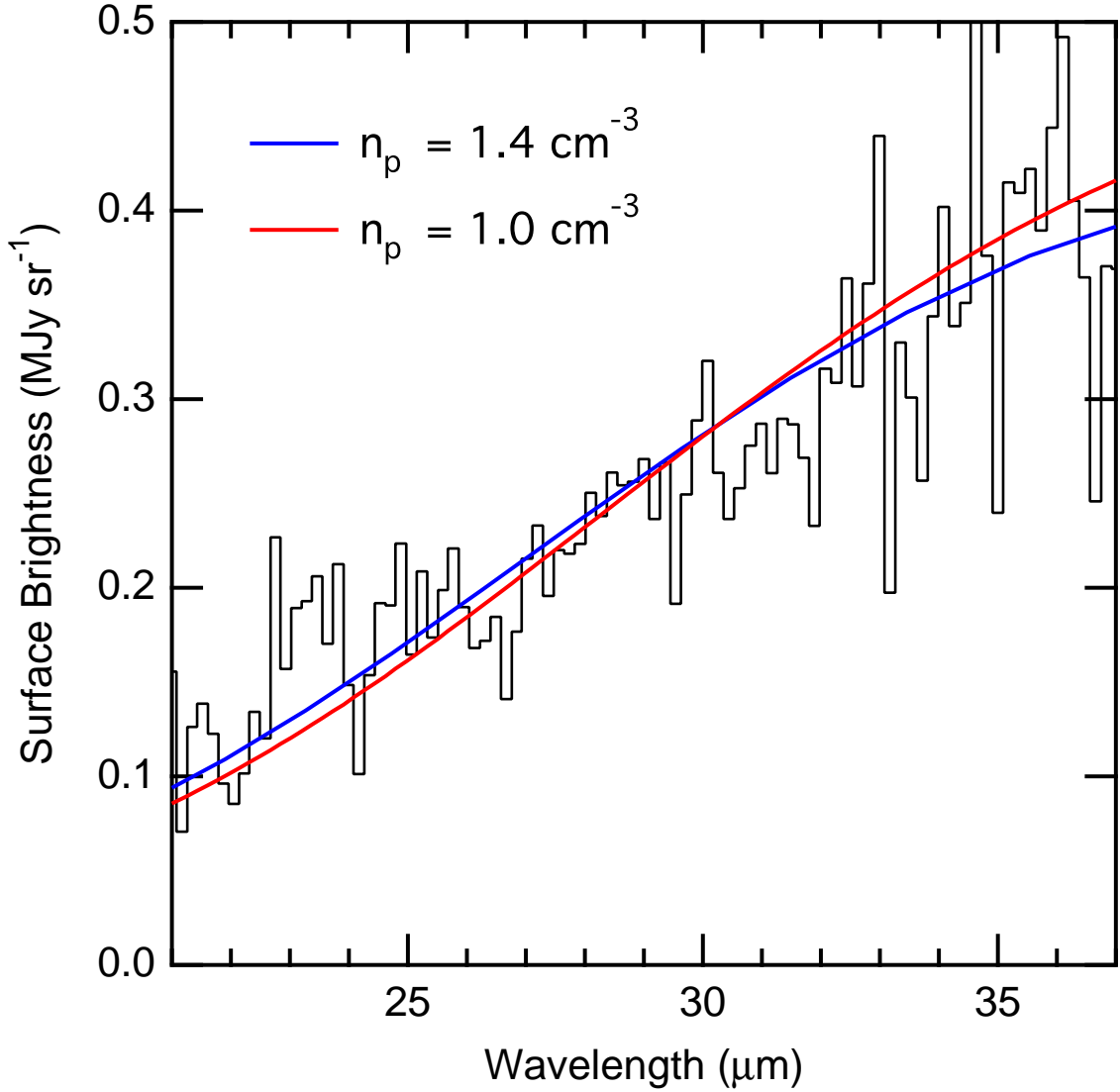}
\caption{IRS spectrum, in black, combining positions 1 and 2 from
  Figure~\ref{irsoverlays}, and containing both orders of the Long-Low
  spectrograph.  The points shown here, at wavelengths $> 21 \mu$m, are all from order 1.  
  The blue curve shows the nominal best-fit model,  and gives a post-shock density in the filament of
  $n_p = 1.4\; {\rm cm^{-3}}$.  The red curve shows a fit with the density
  constrained at $n_p = 1.0\; {\rm cm^{-3}}$, within the limits of  acceptable fits, and consistent with observations in  other bands (text, Section 5.1).  }
\label{irsfit}
\end{figure}

\begin{figure}
\plotone{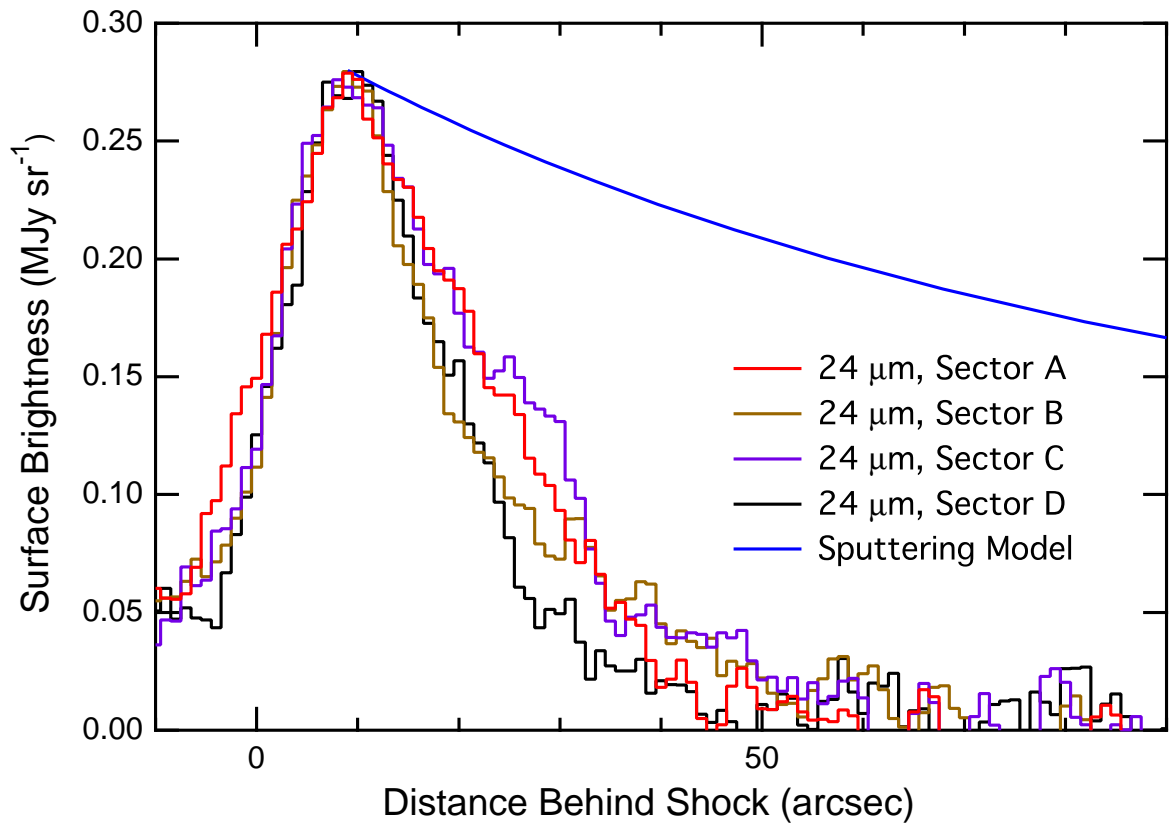}
\caption{A comparison of radial profiles of the 24\,$\mu$m emission (same as Fig.~\ref{profiles}, scaled to a common peak intensity) with a model calculation, which has the post-shock proton density fixed at $n_p = 1\, {\rm cm}^{-3}$, consistent with observations in other bands (text, Section 5.1).  A model assuming expansion into a uniform medium and  normal sputtering rates decays far more slowly than the observations.   
  }
\label{profile_models}
\end{figure}

\end{document}